\documentstyle[twoside,fleqn,espcrc2,psfig]{article}

\newcommand{\AmS}{{\protect\the\textfont2
  A\kern-.1667em\lower.5ex\hbox{M}\kern-.125emS}}

\begin{document}

\title{Unconventional electronic Raman spectra of borocarbide superconductors}

\author{In-Sang Yang$^{1,2}$, M.V. Klein$^2$, T.P. Devereaux$^3$, I.R. Fisher$^4$, and
P.C. Canfield$^4$\\
$^1$Department of Physics, Ewha Womans University, Seoul 120-750, Korea\\
$^2$Science and Technology Center for Superconductivity 
and Department of Physics, University of Illinois, Urbana, IL 61801.\\
$^3$Department of Physics, University of Waterloo, Waterloo, Canada, N2L 3G1\\
$^4$Ames Laboratory, Department of Physics and Astronomy, Iowa State 
University, Ames, IA 50011.\\}       

\begin{abstract}
Borocarbide superconductors, which are thought to be conventional BCS-type superconductors,
are not so conventional in several electronic Raman properties. 
Anisotropic gap-like features and finite scattering strength below the gap were observed 
for the $R$Ni$_2$B$_2$C ($R$ = Lu, Y) systems in our previous study.\cite{yang} 
The effects of Co-doping on the  2$\Delta$ gap-like features and the finite scattering 
strength below and above the gap are studied in 
$R$ = Lu (B = B$^{11}$) system.  
In superconducting states, Co-doping strongly suppresses the 2$\Delta$ peak in 
both B$_{2g}$ and B$_{1g}$ symmetries.
Raman cross-section calculation which includes inelastic scattering shows a relatively 
good fit to the features above the 2$\Delta$ peak, while it does not fully account for the 
features below the peak.

\end{abstract}

\maketitle


Some superconductors that are thought to be of the conventional BCS-type have 
unusual properties. Especially, the borocarbides with the generic formula 
$R$Ni$_2$B$_2$C ($R$ = Y, rare earths) have shown rich physics.\cite{canfield} 
In this article, we further address the peculiar behavior of the 2$\Delta$ peak 
and the sub-gap features, reported in our earlier electronic Raman 
measurements\cite{yang}, in the Lu(Ni$_{1-x}$Co$_x$)$_2$B$_2$C ($x$ = 0.0, 0.015, 0.03). 


The samples measured were single crystals grown by the flux-growth 
method\cite{xu}
and characterized by  temperature-dependence of resistivity   
and magnetization.\cite{resist}
Raman spectra were obtained using a custom-made subtractive triple-grating spectrometer
designed for very small Raman shifts and ultra low intensities.
3 mW of 6471 \AA  \/  Kr-ion laser light was focused onto a spot of
$100 \times 100$ $\mu$m$^2$, in a pseudo-backscattering geometry.
The temperature of the spot on the sample surface  
was estimated to be $\sim$7 K for the superconducting spectra.
The spectra were corrected for the response of the spectrometer and 
the Bose factor. 
Thus they are proportional to the
imaginary part of the Raman susceptibility.


The 7 K Raman spectra in both geometries show 
2$\Delta$-like peak features. 
The intensity of the 2$\Delta$-like peak is stronger and sharper in B$_{2g}$ than in B$_{1g}$.
They are strongly suppressed as pure LuNi$_2$B$_2$C 
is doped by Co on the Ni sites. 
The Co impurities are believed to be nonmagnetic.\cite{resist} 
 
The $T\rightarrow 0$ peak positions of the 2$\Delta$-like feature, 
which were found to 
show BCS-type temperature dependence of the superconducting gap $\Delta$(T), 
are 45 cm$^{-1}$ (B$_{2g}$) and 48 cm$^{-1}$ (B$_{1g}$) for the undoped LuNi$_2$B$_2$C, 
which are much less anisotropic than the values in YNi$_2$B$_2$C (40 and 49 cm$^{-1}$, 
respectively).\cite{yang}
The anisotropy of the peak positions (and thus the apparent gap values) seems to be 
lessened as Co-doping increases. 
The normal state responses (bottom spectra) show 
curvatures at around 35-50 cm$^{-1}$  
depending on the amount of Co-doping, which might be responsible 
for the apparent insensitivity of the 2$\Delta$ peak to the Co-doping 
in the superconducting state. 


A comparision of the data for different Co dopings to the
theory for Raman scattering in disordered conventional
superconductors developed in Ref.\cite{tpd1} is presented in Fig.~1.
The theory has been augmented to include electron-phonon and 
electron-paramagnon inelastic scattering\cite{tpd2}. 
The relevant fit parameters
are the magnitude of gap $\Delta$ and the elastic
scattering rate $1/\tau_{L}$ for $L$ = B$_{1g}$, B$_{2g}$ channels. 
Other parameters entering into the self energies  
(such as the DOS at the Fermi level, Debye energy, electron-phonon coupling,
and the Stoner factor) are taken from Ref. \cite{Pickett}. 
We used $\Delta=20$ cm$^{-1}$ for $x$ = 0 and 0.015
and used $\Delta=19$ cm$^{-1}$ for $x$ = 0.03. 
The values used for $1/\tau_{L}$ 
to fit the $B_{2g} (B_{1g})$ data were 
32, 80, 120 (60, 100, 120) cm$^{-1}$ for $x$ = 0, 0.015, and 0.03, 
respectively. 
The same parameters, except $\Delta=0$, were used to fit the normal state spectra (bottom).
These values are less than a factor of two larger than those
determined by resistivity studies\cite{resist}, 
although these rates are 
necessarily different from the transport rate. 

The theory agrees rather well with the data near the gap 
edge and at higher frequencies, but does not reproduce the spectral weight
observed for small frequency shifts. This intensity might come from
additional bands which have a very small gap\cite{terashima}
or are not superconducting or from nodal quasiparticles.
In the former case, both channels would experience a linear in frequency term coming
from normal scattering processes superimposed on the superconducting response.
In the latter case, the linear rise of the spectra naturally arises from
a gap with line or point nodes, provided the nodes are not coincident
with the nodes of the B$_{1g}$ or B$_{2g}$ vertex.

ISY and MVK were partially supported under 
NSF 9705131 and 9120000. 
ISY was also supported by KOSEF 1999-2-114-005-5.
Ames Laboratory is operated by  U.S. DOE by Iowa State University
under Contract No. W-7405-Eng-82.

\begin{figure}[hbt]
\centerline{%
\psfig{file=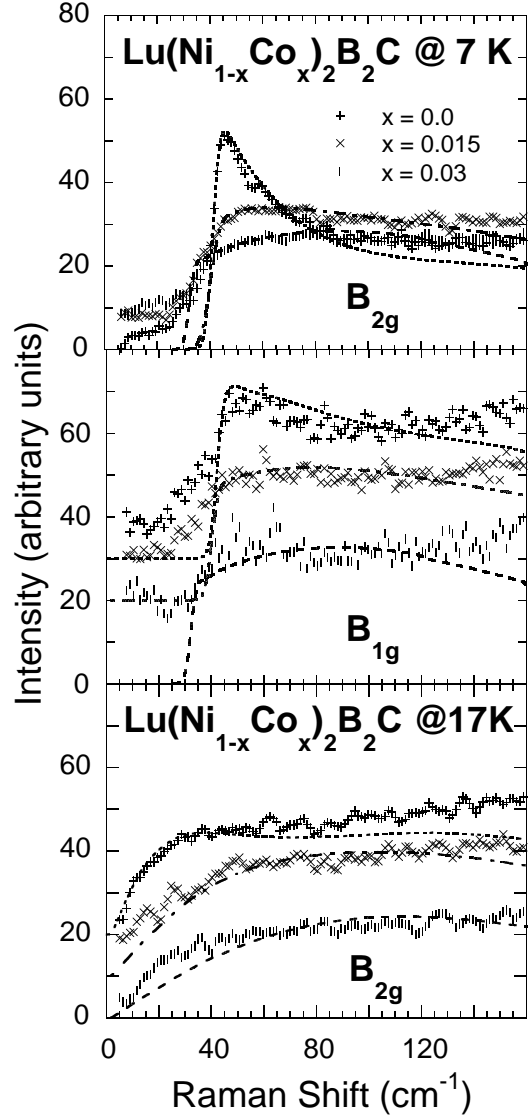}}
\caption{
Raman spectra in B$_{2g}$ (top) and B$_{1g}$ (middle)  from  
Lu(Ni$_{1-x}$Co$_x$)$_2$B$_2$C ($x$ = 0.0, 0.015, 0.03) at 7 K.
The bottom spectra are for B$_{2g}$  
in the normal state (T $\approx$ 17 K), and
similar spectra (not shown) were obtained for  B$_{1g}$. 
Dotted ($x$=0), dot-dashed ($x$=0.015), and dashed lines ($x$=0.03) are 
fit to the theory for conventional superconductors.
The $x$=0 (0.015) data in B$_{1g}$ (at 7 K) and 
B$_{2g}$ (at 17 K)  have been shifted by 30 (20) and
20 (10) units for clarity.
}
\end{figure}


\begin{thebibliography}{9}
\bibitem{yang} In-Sang Yang {\it et al.\/}, 
cond-mat9910087 (1999).
\bibitem{canfield}
See, for example, 
P. C. Canfield   {\it et al.\/},
Physics Today {\bf 51}, 40 (1998).
\bibitem{xu}
M. Xu  {\it et al.\/}, 
Physica C {\bf 227}, 321 (1994).
\bibitem{resist}
K. O. Cheon {\it et al.}, Phys. Rev. B {\bf 58}, 6463 (1998).
\bibitem{tpd1} T. P. Devereaux, Phys. Rev. B {\bf 45},
12965 (1992).
\bibitem{tpd2} T. P. Devereaux and D. Belitz, Phys. Rev. B {\bf 44},
4587 (1991).
\bibitem{Pickett}
W. Pickett and D. J. Singh, Phys. Rev. Lett. {\bf 72}, 3702 (1994).
\bibitem{terashima}
T. Terashima {\it et al.}, Phys. Rev. B {\bf 56}, 5120 (1997).
\end{thebibliography}
\end{document}